\documentclass[%
 aip,
 rsi,
 amsmath,amssymb,noeprint,
 reprint,%
 numerical,
]{revtex4-2}
\usepackage{graphicx}
\usepackage{dcolumn}
\usepackage{bm}
\usepackage{amsmath}

\usepackage[utf8]{inputenc}
\usepackage[T1]{fontenc}
\usepackage{mathptmx}
\usepackage{hyperref}
\usepackage[all]{hypcap}
\usepackage[usenames,dvipsnames]{color}
\hypersetup{urlcolor=Blue, citecolor=Blue,linkcolor=Blue,colorlinks=true}

\usepackage{titlesec}
\titlespacing*{\section}{0pt}{0.5\baselineskip}{0.5\baselineskip}
\titlespacing*{\subsection}{0pt}{0.5\baselineskip}{0.5\baselineskip}
\titlespacing*{\subsubsection}{0pt}{0.5\baselineskip}{0\baselineskip}
\newcommand{\sref}[2]{\hyperref[#1]{Fig. \ref{#1}#2}}

\newcommand{\comment}[1]{}

\begin{document}


\title{An Imaging Refractometer for Density Fluctuation Measurements in High Energy Density Plasmas}

\author{J. D. Hare}
\email[Current Address: Plasma Science and Fusion Center, Massachusetts Institute of Technology, Cambridge, MA, USA. ]{jdhare@mit.edu}
 \affiliation{Blackett Laboratory, Imperial College, London, SW7 2AZ, UK}
\author{G. C. Burdiak}%
 \affiliation{First Light Fusion Ltd, 10 Oxford Industrial Park, Yarnton, Kidlington OX5 1QU, UK}
\author{S. Merlini}
\author{J. P. Chittenden}
\author{T. Clayson}
\author{A. J. Crilly}
\author{J. W. D. Halliday}
\author{D. R. Russell}
\author{R. A. Smith}
\author{N. Stuart}
\author{L. G. Suttle}
\author{S. V. Lebedev}
\email[Corresponding author: ]{s.lebedev@imperial.ac.uk}
 \affiliation{Blackett Laboratory, Imperial College, London, SW7 2AZ, UK}

\date{\today}

\begin{abstract}
We report on a recently developed laser-probing diagnostic which allows direct measurements of ray-deflection angles in one axis, whilst retaining imaging capabilities in the other axis.
This allows us to measure the spectrum of angular deflections from a laser beam which passes though a turbulent high-energy-density plasma.
This spectrum contains information about the density fluctuations within the plasma, which deflect the probing laser over a range of angles.
We create synthetic diagnostics using ray-tracing to compare this new diagnostic with standard shadowgraphy and schlieren imaging approaches, which demonstrates the enhanced sensitivity of this new diagnostic over standard techniques.
We present experimental data from turbulence behind a reverse shock in a plasma and demonstrate that this technique can measure angular deflections between 0.06 and 34 mrad, corresponding to a dynamic range of over 500.
\end{abstract}

\maketitle

\section{Introduction\label{sec:intro}}

Turbulence drives inhomogeneities in fluids, creating a cascade of fluctuations from the driving scale down to the scale at which viscous dissipation dominates.
These fluctuations create density gradients over a wide range of scales, which deflect a probing laser beam.
These deflections have attracted widespread interest in aero-optics\cite{Sutton1969} and plasma physics.\cite{Mazzucato1966, Kasim2017, White2019, Collins2020}

Two techniques for measuring these density fluctuations are often used in high-energy-density (HED) plasmas: shadowgraphy and schlieren imaging.
In shadowgraphy, density fluctuations act as natural lenses, deflecting rays, and producing intensity variations (light and dark regions) in the probing laser.
This natural focusing removes the one-to-one relation between the image on the detector and the object, and so it is not possible to measure spatial scales directly from a shadowgram.\cite{Settles2001, Nye1999} 

In schlieren imaging, a stop is placed at a focal plane to cut off rays with specific deflection angles. 
With a finite sized source, the schlieren image contains information on the gradients sampled by the probe beam, but with a laser source the measurement is often only sensitive to a very narrow range of gradients.
Therefore laser schlieren images are almost binary, showing the location of density gradients but not their magnitude.\cite{Settles2001}
With both shadowgraphy and schlieren, the analysis of turbulence is often similar: the image is Fourier transformed, the spectrum of intensity variations is linked to the spectrum of density fluctuations, and a power-law fit is made and compared with theoretical predictions.\cite{White2019, Collins2020}

These imaging techniques have two significant drawbacks.
First, large density gradients result in ray crossings, leading to bright regions or ``caustics''\cite{Settles2001, Nye1999} which prevent a unique reconstruction of the density fluctuations.
Second, the smallest resolvable length scale is at least a few times the detector pixel size, so that in a realistic setup the difference between the driving length scale and the smallest resolvable length scale is often only a single order of magnitude.
It is difficult to fit a power-law spectra over such a limited range,\cite{Goldstein2004b} and it is difficult to probe the full inertial range down to the dissipation scale.

The spectrum of deflection angles within a probing laser beam is directly related to the spectrum of density fluctuations within the turbulent plasma.
This deflection angle spectrum can be simply measured by placing a detector at the focal plane of a lens, which contains a Fourier transform of the probing beam, which is the two-dimensional spectrum of deflection angles.\cite{hecht2012optics}
However, this spectrum lacks information about spatial variations in the properties of the turbulence.

\begin{figure}[!h]
	\label{fig:precursor}
	\includegraphics[scale=1]{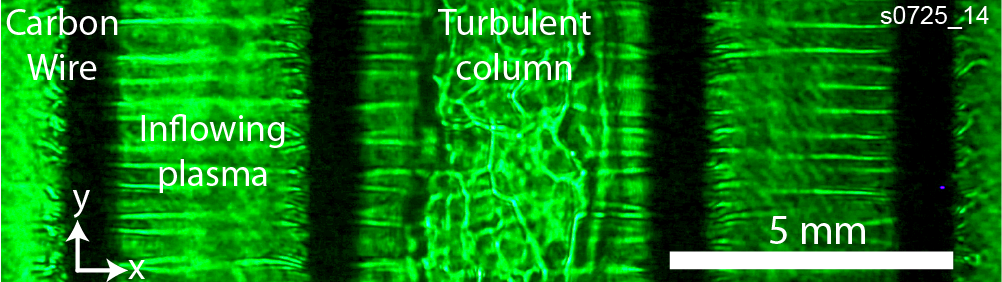}
	\centering
	\caption{A shadowgraphy image of a turbulent plasma column formed at the centre of an imploding, eight-wire, carbon z-pinch, as discussed in Ref. \onlinecite{Hare2017a}.}
\end{figure}

In this paper we present a new hybrid diagnostic, an imaging refractometer, which has spatial resolution along one axis and angular resolution along the perpendicular axis.
This is a powerful technique when the properties of turbulence within an HED plasma are homogeneous in one direction, eg. an experiment with cylindrical or planar geometry.
In this case, integrating over this spatial dimension can provide angular resolution with an increased signal-to-noise ratio.
For example, in \sref{fig:precursor}{} the turbulence in a plasma column formed inside a wire array z-pinch appears homogeneous in $y$, but varies in $x$.

This new diagnostic uses physical optics to image a one-dimensional Fourier transform of a probing beam.
This diagnostic's dynamic range is over 500, far in excess of techniques using digital Fourier transforms, and is not limited by the formation of caustics.
We give an example from an experiment in which density fluctuations form behind a reverse shock.

\begin{figure*}[!t]
	\label{fig:optics}
	\includegraphics[scale=1]{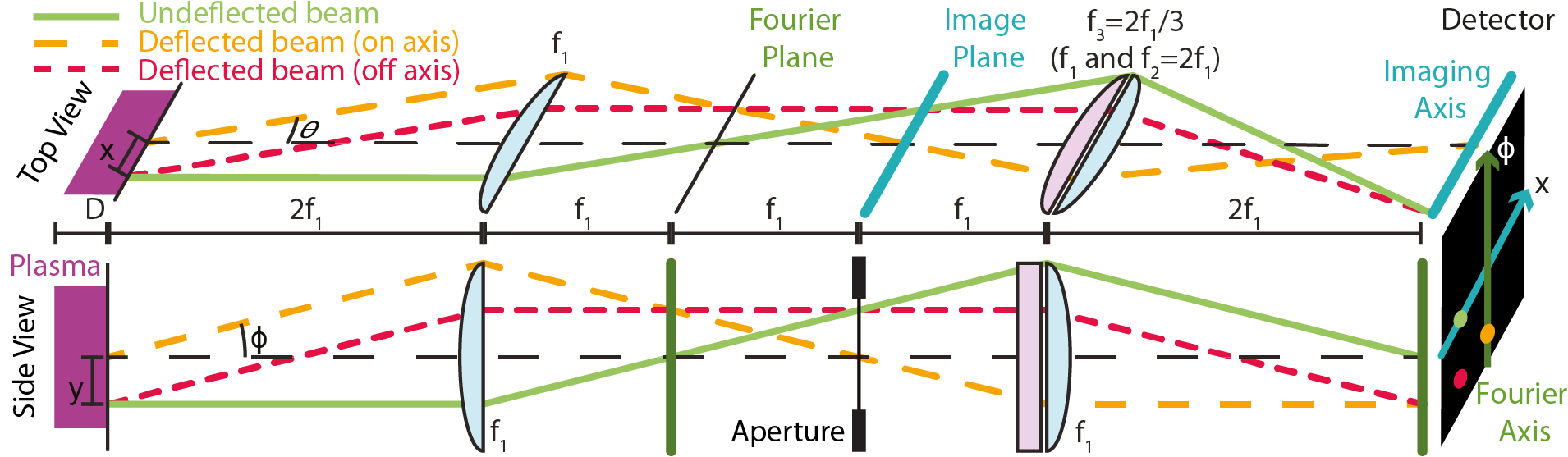}
	\centering
	\caption{A ray diagram of the imaging refractometer, from two orthogonal directions. An initially collimated laser propagates through a plasma of length D from the left. The first lens (cyan) forms a Fourier plane and an image plane, which are imaged by a composite optic consisting of a spherical lens (cyan) and a cylindrical lens (pale purple). This results in a composite image, with b) one axis of spatial resolution, and a) one axis of angular resolution. The diagram shows three example rays: green; a collimated ray, and red and orange; deflected, collimated rays.}
\end{figure*}

\section{Optical configuration\label{sec:optics}}
We will illustrate the operation of this diagnostic for the specific focal length relations shown
in \sref{fig:optics}{}, with a nearly collimated laser beam travelling from left to right, which passes through a turbulent plasma of length $D$.
After the first lens (spherical, focal length $f_1$, cyan) there are two planes of interest, the image plane and the Fourier plane.
At the image plane, there is an image of the probing laser beam at the exit of the plasma, containing the spatial distribution of the rays.
At the Fourier plane, there is the Fourier transform of the probing laser beam, containing the angular distribution of the rays.

The second optic consists of a spherical (cyan) lens and a cylindrical (pale purple) lens placed next to each other, which focuses light differently along two orthogonal axes.
In the Imaging axis (\sref{fig:optics}{a}), the combination of the cylindrical and spherical lenses focus the rays with $f_3=2f_1/3$, reproducing the image plane at the detector.
In the Fourier axis (\sref{fig:optics}{b}) the spherical lens focuses the rays with focal length $f_1$, forming an image of the Fourier plane at the detector.
Hence the x-axis (y-axis) of the image at the detector shows the spatial (angular) distribution of rays at the object plane.

In \sref{fig:optics}, the green line represents a collimated beam propagating parallel to the optical axis. This ray crosses the optical axis at the Fourier Plane, and crosses the image plane at the same distance from the optical axis as it originated.
After the second lens, this ray arrives at the optical axis in the Fourier direction ($\phi=0$), and in the Imaging axis it arrives at the detector at twice its initial distance from the optical axis, a magnification of $M=2$.
The orange and red rays represent a collimated laser beam deflected by an angle with respect to the optical axis. They cross the first Fourier plane off-axis, and reach the image plane with $M=1$.
The second lens images both rays to the same off-axis location in the Fourier axis, indicating that the ray exited the plasma with $\phi\ne0$.
The rays are imaged with $M=2$ onto the Imaging axis.

In the paraxial approximation this optical system can be treated using standard Ray-Transfer Matrix techniques.\cite{siegman1986lasers}
We treat all of the lenses as thin, and calculate the position of a ray $X_f$ at the detector as:
\begin{align}
X_0 =\left[
\begin{matrix}
x\\
\theta\\
y\\
\phi
\end{matrix}
\right]
\rightarrow
X_f =\left[
\begin{matrix}
2 x\\
\theta/2 + 2 x/f_1\\
- f_1 \phi\\
 y/f_1
\end{matrix}
\right]
\label{eqn:rtm}
\end{align}
where $X_0$ is the ray at the exit of the plasma, and for $x$, $y$, $\theta$ and $\phi$ see \sref{fig:optics}{}.
The x-direction is imaging with a magnification of 2, and has no dependence on the initial angle.
The y-direction only depends on the angle of the initial ray ($\phi$) from the z axis in the y-z plane.
The y location of the ray at the detector is $f_1\phi$, which defines the angular sensitivity of the detector.
The resolution of the instrument is set by any imperfections in the optics, the initial divergence of the beam, the pixel size of the detector, and limitations due to diffraction.

\section{Implementation for HED experiments\label{sec:setup}}

In an inhomogeneous medium, rays are deflected by refractive index gradients normal to the direction of propagation:
\begin{align}\alpha = \int \nabla N dl=\int \frac{1}{2}\frac{\nabla n_e}{n_{cr}} dl,\label{eqn:alpha}\end{align}
where $\alpha$ is in radians and the gradient is taken perpendicular to the path $dl$.\cite{Hutchinson2002}
In a plasma the refractive index ($N$) is related to the electron density $n_e$ and the critical density $n_{cr}=\omega^2\epsilon_0 m_e/e^2 = 1.12 \times 10^{21} \lambda_0^{-2}$, where $\lambda_0$ is the freespace wavelength in $\mu$m, and $n_{cr}$ is in units of cm$^{-3}$.
It is these deflections which are measured by the imaging refractometer.

In order to measure the deflection angle with a high dynamic range, we require a laser which a) has a small angular divergence, b) has a smooth beam profile, c) is highly reproducible, d) has a short coherence length, and e) is sufficiently intense to overcome the self-emission of the plasma.

The smoothness of the laser beam ensures that any variations in intensity are due to deflected light, rather than variations in the initial source profile. 
The high collimation ensures that the incoming laser beam can be well described by a single Fourier component, and hence focuses to a very narrow line, which contributes to the resolution of this diagnostic. 
The high reproducibility is necessary to compare the results obtained in the absence of the plasma and with the plasma, to determine how the intensity of the laser beam is redistributed.
The short coherence length is necessary to minimise interference effects which degrade laser-based schlieren and shadowgraphy techniques. \cite{Settles2001}
These can cause variations in the measured intensity which could be mistaken for fluctuations in the measured deflection angle.
The intensity of the laser must be much larger than the measurable self-emission of the plasma, which degrades the signal-to-noise ratio.

For our experiments we use the long-pulse arm of the Nd:Glass CERBERUS laser (1053 nm, 100 mJ, 1 ns).\cite{Swadling2014a}
This laser passes through several vacuum spatial filters, resulting in a divergence $<0.05$ mrad, and the reproducibility of the intensity profile is better than $1\%$.
The laser has a low coherence length ($<1$ mm), which minimises interference effects.
Using an infra-red beam also enhances the sensitivity of the system over visible light, as $\alpha\propto1/n_{cr}\propto\lambda^2$.

We implemented this diagnostic with $f_1=200$ mm (see \sref{fig:optics}), which is determined by the radius of our vacuum chamber and the available optics.
Therefore $f_1=200$ mm for the two spherical lenses (50 mm diameter Thorlabs Achromatic doublet AC508-200-C-ML), and the cylindrical lens has $f_2=400$ mm (60 mm $\times$ 30 mm Thorlabs plano-convex LJ1363L2-C).
The second optic consists of the cylindrical lens and spherical lens placed back to back, with $f_3=400/3$ mm.
A 50:50 beam splitter (50 mm diameter Thorlabs UV-Fused silica BSW30) immediately after the first lens directs half of the probe light into a camera at the image plane, resulting in a 2D shadowgraphy image of the plasma object.
The optics are coated to suppress reflections at the laser wavelength, $\lambda_0 = 1053$ nm, and the beam diameter is $\approx 22$ mm.

For the shadowgraphy arm, we use a 15 MP DSLR (Canon 500D), with the IR filter removed and no lens mounted to the body.
All the optics are mounted to a breadboard outside the vacuum chamber.
The exposure time of the camera is 1.3 s, so the effective exposure time of the image is instead set to \mbox{1 ns} by the duration of the laser pulse.

For the imaging refractometer arm, we use an ATIK 383L+ camera, which has a cooled 8 MP ($3448 \times 2574$ pixels, $18 \times 13.5$ mm), monochrome Kodak KAF-8300 CCD with 16-bit depth.
The high bit-depth and exceptionally low noise of this CCD is vital to making measurements of the deflected rays with a high dynamic range.
We use a 1 s exposure, with the \mbox{1 ns} laser pulse setting the temporal resolution.

To ensure that the properties of the turbulence are homogeneous in the vertical direction, we introduce an aperture at the image plane after the first lens, which blocks rays from outside our region of interest.
This ensures that we only measured the spectrum of angles for rays which propagate through similar regions of plasma.
We note that with a very narrow, slit-like aperture, our diagnostic resembles that in ref. \onlinecite[p.~313]{Ascoli-Bartoli1965}.

\begin{figure*}[!th]
	\label{fig:comparison}
	\includegraphics[scale=1]{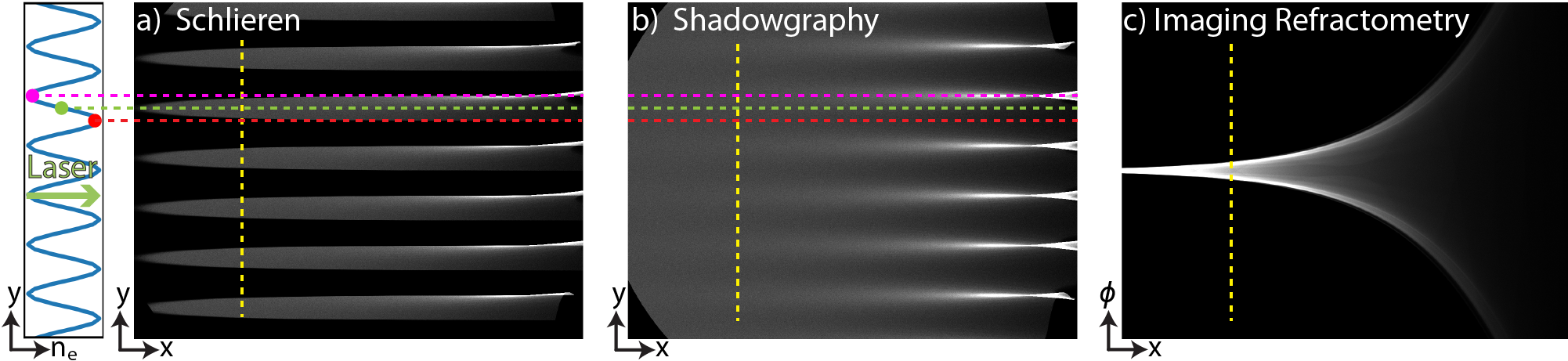}
	\centering
	\caption{Synthetic images for a) Schlieren, b) Shadowgraphy and c) Imaging Refractometry, for $10^7$ rays passing through a sinusoidal density perturbation in $y$ (shown on the left), which increase in strength in $x$. The locations of minima, steepest gradient and maxima in density are shown using red, green and purple lines, respectively. The ray locations at the detector binned into $3448\times2574$ pixels occupying $18\times13.5$ mm. The angular range of the imaging refractometer is $\phi=\pm34$ mrad.}
\end{figure*}
\section{Comparison with shadowgraphy and schlieren}

To compare this system with existing laser-imaging systems, we used a custom ray tracing code to create synthetic diagnostics representing the imaging refractometer, shadowgraphy and schlieren imaging systems.
The imaging refractometer is as shown in \sref{fig:optics}, and the shadowgraphy and schlieren systems use the optics shown in \sref{fig:optics}{a}.
For the schlieren system, a knife edge is placed at the Fourier plane, $f_1$ after the first lens.
The knife edge is slightly displaced to form a dark-field system, which images only rays deflected by $\phi>0.1$ mrad.

In a real system, each optic acts as a finite size aperture, resulting in vignetting and related effects which are captured in our custom-written ray transfer matrix solver by discarding rays which fall outside of any aperture.
We approximate the lenses as perfect, without any aberrations, which is a good approximation for the achromatic lenses (which also correct spherical aberrations), but in reality the cylindrical lens will cause aberrations in the imaging axis.
The cylindrical lens changes the path of the rays for the imaging axis only, and so these aberrations do not degrade the angular resolution.

As a test problem, we consider a sinusoidal perturbation to a plasma, with an amplitude that grows in a transverse direction. 
The plasma is uniform along the direction of laser propagation ($z$). 
We generate a density field described by $n_e(x,y) =n_{e0} 10^{x/s}[1+\cos{\left(2\pi y/L_y\right)}]$, with periodic perturbations in $y$ which grow exponentially in $x$.
We use $n_{e0} = 2\times 10^{17}$ cm$^{-3}$, $s=4$ mm and $L_y=1$ mm, and the plasma is modelled as a cube with 101 grid points per axis, and a side length of 10 mm ($x,y,z=[-5,5]$ mm). 

We generate 9.6$\times10^8$ test rays randomly located within a beam of diameter 10 mm.
The initial ray angles are drawn from a normal distribution with width $0.05$ mrad, corresponding to the experimentally measured response function.
This accounts for the imperfect collimation of the input beam, and any optical aberrations.
These rays are traced through the density cube using a ray tracer, which interpolates the refractive index gradient at the ray location.\cite{Kaiser2000}
After the rays have exited the plasma, they are propagated through the three optical systems using a ray transfer matrix technique.

The results are shown in \sref{fig:comparison}{}.
The schlieren image (\sref{fig:comparison}{a}) shows a series of bright horizontal bands, centred on the maximum value of $\partial n_e/\partial y$, which increase in width from left to right, with caustics forming at the far-right of the image.
These bands are where the gradients deflect rays in the $+\phi$ direction.
The schlieren effect results in a binary image, because laser-based schlieren imaging is only sensitive to a narrow range of density gradients - for larger density gradients, all of the rays either pass the knife edge, or they are blocked.\cite{Settles2001}

Intensity variations in laser-schlieren imaging are often due to shadowgraphy effects (proportional to the second derivative of the electron density) rather than schlieren effects (proportional to the first derivative).
This is particularly clear inside the caustic forming region on the right of the schlieren image, which is identical to the intensity profile in the shadowgraphy image.
Therefore digital Fourier transforms of schlieren images (such as in Ref.  \onlinecite{Collins2020}) may be dominated by shadowgraphy effects, which affects the interpretation of the power spectrum and how it relates to the spectrum of density fluctuations.

The synthetic shadowgraphy diagnostic (\sref{fig:comparison}{b}) shows horizontal bands in the same locations as the schlieren image, but the dynamic range is not binary.
The contrast of the bands increases as the density gradients increase from left to right.
Towards the right of the image, the sinusoidal perturbations focus the rays, causing bright central bands surrounded by darker bands.
At the far-right of the image, the bright bands broaden again as the rays cross, and caustics form.

The imaging refractometer (\sref{fig:comparison}{c}) shows a bright bounding curve which increases exponentially from $\phi=0$, reflecting the exponential density ramp from left to right.
This bright curve represents the regions of highest deflection, or steepest electron density gradient, and bounds a region of lower intensity, corresponding to smaller deflection angles.

By considering the vertical dashed yellow lines in \sref{fig:comparison}{}, it is clear that the imaging refractometer is more sensitive to angular deflections than shadowgraphy and schlieren techniques.
At this location, the schlieren image shows only the location of the density gradients and the shadowgraphy shows very minimal variations in intensity, but the imaging refractometer records deflections of 1.8 mrad, 30 times large than its 0.06 mrad resolution.
Indeed, the schlieren technique is really only appropriate for showing the location of steep density gradients, and does not provide any information about their properties.
The shadowgraphy technique does provide more information, but in practice the double-integration necessary to retrieve the properties of the density perturbation is very sensitive to experimental noise.
Other statistical techniques may be more appropriate for retrieving the density modulations, providing the input laser beam is well characterised. \cite{Kasim2017}
The downside of the imaging refractometer is that it gives no information about the location of the density perturbations in the $y$ direction, and so is ideally complemented by the inline shadowgraphy arm we discussed earlier.

\section{Experimental results\label{sec:results}}

\begin{figure}[h]
	\label{fig:experiments}
	\includegraphics[scale=1]{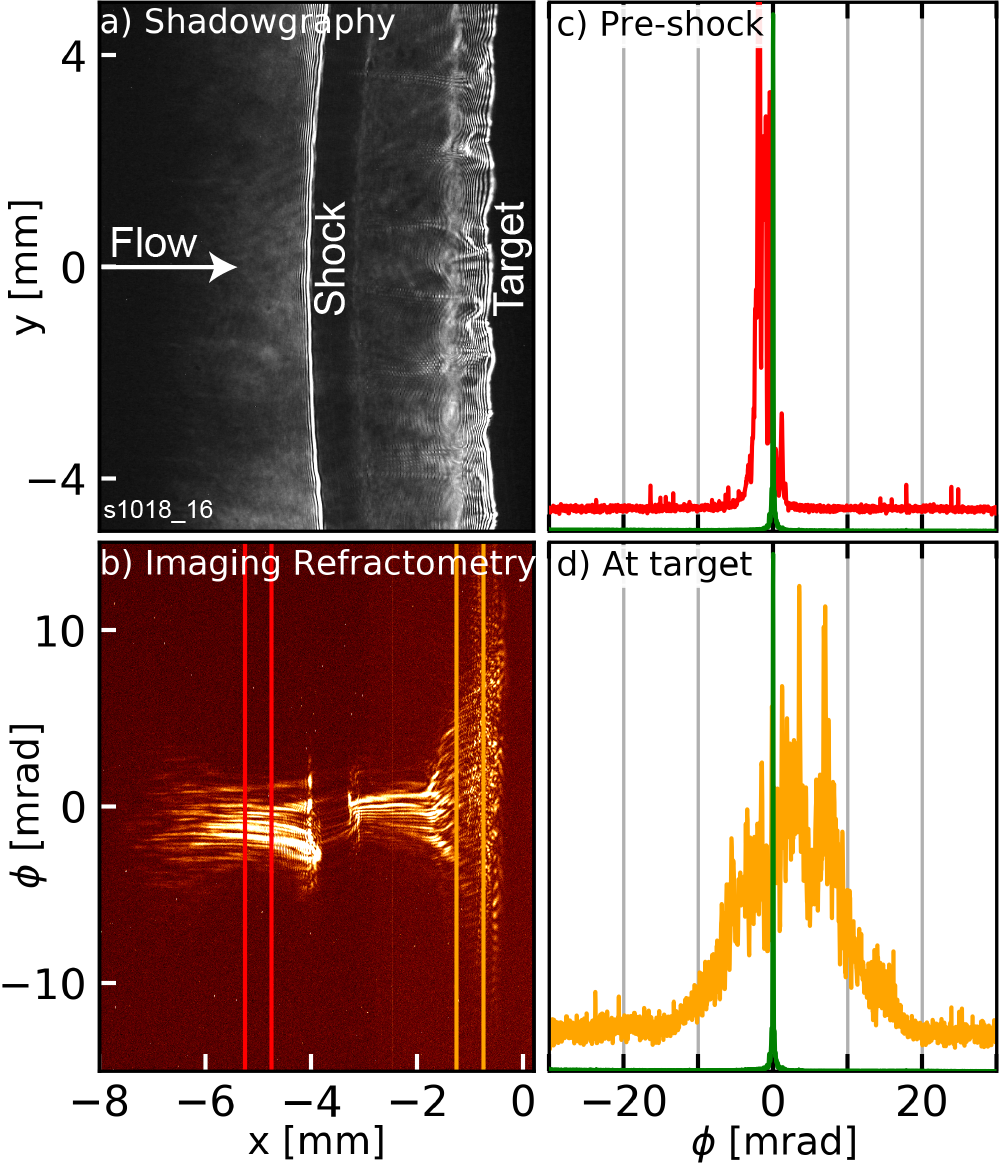}
	\centering
	\caption{Experimental data from a pulsed-power driven supersonic plasma colliding with a planar target. Images are shown from a) shadowgraphy and b) the imaging refractometer. The orange and red boxes show the positions of the lineouts from the refractometry in two locations, c) pre-shock and d) at the target, with the response function in green. The y-axis in c) and d) is scaled to show the shape of each lineout, rather than absolute intensity.}
\end{figure}

To demonstrate the capabilities of this diagnostic, we carried out an experiment on the MAGPIE pulsed-power generator (1.4 MA peak current, 250 ns rise time) \cite{Mitchell1996}.
We used an exploding wire array \cite{Harvey-Thompson2009, Lebedev2014} to produce a super-sonic, magnetised aluminium plasma outflow.
In the shadowgraphy image in \sref{fig:experiments}{a}, the flow propagated from left to right and the target was on the right.
This outflow collided with a planar target and created a reverse shock, which propagated back (left) towards the array.
The shock is very clear, an intense narrow band 0.06 mm wide, which represents a strong, focusing density gradient.
The intensity variations are small in the upstream and immediate downstream flow, but close to the planar target there are small scale intensity fluctuations.

We can gain further insight from \sref{fig:experiments}{b}, which shows data from the imaging refractometer.
Here we see that the upstream flow is structured, with horizontal lines indicating distinct deflection angles.
This is due to the well-understood axial-modulation of the ablation in exploding and imploding wire arrays, also seen in \sref{fig:precursor}{}.\cite{Chittenden2008a}
Immediately post-shock the imaging refractometer records no intensity due to the strong shadowgraphy effect which deflects the rays horizontally.
Further right, there is a region with smaller deflection angles than in the upstream region, suggesting that the axial modulations are damped after the shock.
In this region, the shadowgraphy shows no intensity variations, but they are clearly visible in the imaging refractometer, indicating the higher sensitivity of this new diagnostic to small deflections.
Close to the target we observe an abrupt transition, to a wide spread of deflection angles, corresponding to a region with density fluctuations over a broad range of spatial scales, which may indicate turbulence.

The spatial variations in the deflection angle spectrum are apparent in \sref{fig:experiments}{c and d}, which shows two lineouts from \sref{fig:experiments}{b} (averaged over 0.5 mm in $x$), as well as the detector response functions measured in a shot without plasma, which also determines the location of $\phi = 0$ .
The response function, shown in green in \sref{fig:experiments}{c and d}, is well approximated by a normal distribution with a width of 0.06 mrad, which is over 500 times smaller than the maximum angle of 34 mrad, which is determined by the size of the detector compared with $f_1$ through eqn. \ref{eqn:rtm}.
This response function is set by the initial quality of the probing laser beam, the optics, diffraction effects and the resolution of the detector.

The pre-shock spectrum is significantly broader than the response function, consistent with the axial modulation of the incoming flow as discussed above.
The spectrum at the target is much broader still, by a factor of 200 greater than the response function, which again provides evidence for a broad range of density fluctuations consistent with turbulence.

\section{Conclusions}
In this paper we have outlined the theory and application of a new hybrid diagnostic, which provides very high angular and spatial resolution (0.06 mrad and 0.06 mm).
When used with a short optical pulse, we also achieve high temporal resolution.
We explored the theory of this diagnostic using a ray-transfer matrix based approach, and we used a ray-tracing technique to compare this imaging refractometer with the familiar shadowgraphy and schlieren imaging diagnostics.
The spectrum of deflection angles is directly linked to the spectrum of density fluctuations, and the imaging refractometer directly measures this spectrum of angular deflections in one direction.
In contrast, schlieren and shadowgraphy imaging require the spectrum to be inferred from digital Fourier transforms, which suffer from limited resolution, and are strongly affected by ray crossing which form caustics.
Neither limitation applies to the imaging refractometer, which offers a new way to study turbulence in HED plasmas.

We demonstrated the real-world capabilities of the imaging refractometer in a pulsed-power driven high-energy-density plasma experiment, using a supersonic outflow from an exploding wire array to create turbulence behind a reverse shock, which we simultaneously imaged using shadowgraphy and the imaging refractometer.
In future, we will use ray-tracing to calculate the propagation of a laser beam through a turbulent medium with a given spectral index and intermittency properties.
The output rays will then be transferred through our optical system, so that we can link the measured spectrum of angular deflections directly to the spectrum of density fluctuations within a turbulent plasma.

\section*{Acknowledgements}

This work was supported in part by the Engineering and Physical Sciences Research Council (EPSRC) Grant No. EP/N013379/1, and by the U.S. Department of Energy (DOE) Awards DE-F03-02NA00057, DE-SC-0001063 and DE-SC0020434.
The data supporting this study are available from the corresponding author upon reasonable request.

\bibliography{library}

\end{document}